# Generalized Friedel's sum rule and anomalies of the mobility due to the resonant scattering of electrons by donor impurities in semiconductors

V.I. Okulov

Institute of Metal Physics Urals Division of RAS, Ekaterinburg

On the basis of Friedel's approach the theoretical description of the effects of resonance scattering of conduction electrons by donor impurities in semiconductors with allowance for the stabilization of electron concentration in coinciding the Fermi energy with the resonance level energy has been developed. It has been shown that such a stabilization gives rise to the appearance of maximum in concentration dependence and to its related anomalies of temperature dependences of electron mobility. The advantage of the application of the approach based on the proposed theory to the interpretation of experimental data on mercury selenide with iron impurities is discussed.

Introduction

In quantum-mechanical scattering theory of an electron by static centre the resonance scattering at which the energy of the electron under scattering is near the energy of resonant level belonging to continuous spectrum has been studied in detail. If the asymptotic value of the wave function $\psi(r)$ of an electron far from the centre is described by typical expression:

$$\psi(\mathbf{r}) \sim e^{i\mathbf{kr}} + f\, e^{ikr}/r, \qquad (1)$$

the corresponding behavior of scattering amplitude $f$ depending on the wave vector $k$ or energy $\varepsilon = \hbar^2 k^2/2m$ is resonant. In asymptotic values of the terms $\psi_l(r)$ of the wave function (1), corresponding to the definite values of orbital moment,

$$\psi_l(r) \sim (1/r)\exp(i\delta_l)\sin(kr - \pi l/2 + \delta_l), \qquad (2)$$

the resonance manifests itself in specific dependence of the derivative with respect to energy of certain phase shift $\delta_r$ from the set $\delta_l$ :

$$d\delta_r/d\varepsilon = \Delta/[(\varepsilon - \varepsilon_r)^2 + \Delta^2] \qquad (3)$$

Eqn. (3) is valid for the energies $\varepsilon$, close to the resonant $\varepsilon_r$, thereby the width of resonance level $\Delta$ is considerably smaller than $\varepsilon_r$. In the limit of very small $\Delta$ the resonant state has almost the same properties as bound state, however, the form of asymptotic value (1) including the contribution of free motion is of principal. The wave function of bound state $\psi_c(r)$ is exponential in character:

$$\psi_b(r) \sim \exp(-\gamma r), \qquad (4)$$

where $\gamma$ the is the real positive quantity. The difference of the wave function (4) from (1) can be a factor. In scattering the conduction electrons by impurity centers in solid states the resonance may occur in the case, when the impurities are donor ones and donor level falls within the conduction band. The theory of these effects in metals has been developed by Friedel [1]. The specific behaviors similar to those which were studied by Friedel can arise also in low-temperature properties of semiconductors, in particular, in the case of doping the semiconductor by the impurities of transition elements. However, in semiconductor physics the situations, when impurity donor level lies in forbidden band and conforms to bound state or the temperature is so high that the difference between the bound and resonance states can be negligible, are more widely known and more intensively studied. Besides the theories developed are often concerned with the range of high concentrations of impurities when the impurity bands are formed by donor levels. Therefore the established approaches to the interpretation of the scattering by resonant donor levels lying in conduction band did not take into account, as a rule, the above difference of principal existing between resonant and bound states at low concentration of impurities. Thereby such a difference was reduced to purely quantitative relation of the widths of levels. It should be noted that the application of such approaches can result in non-adequate description of ground state of the system of electrons and donor impurities interacting at resonance.

In addition the most interesting effects applied to semiconductors brought out by Friedel on the base of known sum rule for phase shifts prove to be not revealed as well. The effects above are concerned with non-monotonic behavior of conductivity against the concentration of impurities. The aim of the present report is to develop consistent description of low-temperature effects of resonant electron scattering by donor impurities in semiconductors by applying the concepts of scattering theory and Friedel's approach. It will be shown how the known phenomenon of the stabilization of electron concentration gives rise to concentration maximum of electron mobility and specific behavior of the dependences of mobility on the temperature. In conclusion we shall discuss the possibilities of applying the results for interpretation of experimental data and give critique of the approaches having been used for this purpose up to now.

1. Resonance phase and sum rule for the scattering phases

For describing the energy spectrum of electrons and scattering potential of impurities let us use simple isotropic models. Let the impurity concentration $n_i$ be small enough so as at a distance $r = (3/4\pi n_i)^{1/3}$, equal to the radius of sphere falling at one impurity, the wave function of an electron is described by the asymptotic values (1), (2). Then the expression for $n(\varepsilon)$ obtained by Friedel [1] is valid: ($n(\varepsilon)$ is the averaged over volume density of electrons occupying the states with the energies from 0 to the given value $\varepsilon$)

$$n(\varepsilon) = n_e(\varepsilon) + n_i z(\varepsilon) \qquad (5)$$

In this expression $n_e(\varepsilon)=k^3(\varepsilon)/3\pi^2$, the concentration of free electrons obtaining in averaging the first term in Eq.(1); the limiting wave vector $k(\varepsilon)$ can be determined from the equality $\varepsilon = \hbar^2 k^2(\varepsilon)/2m$. The second term in Eq. (5) is the contribution to $n(\varepsilon)$ of inhomogeneous part of electron density localized near the impurity centers because of the scattering. Relative part $z(\varepsilon)$ of this contribution falling at one scattering center is expressed in terms of scattering phases $\delta_l$, entering (2):

$$z(\varepsilon) = (1/\pi)\sum_l v_l \delta_l(\varepsilon) = (1/\pi) \sum_l 2(2l+1) \delta_l(\varepsilon) \qquad (6)$$

Here partial contributions to $z(\varepsilon)$ proportional to the scattering phases from all the electron states are summarized. Accordingly the quantity $v_l = 2(2l+1)$ is the multiplicity of degeneracy of the state with the given $l$.

Considering the applications of the Eqs. (5), (6) to the scattering of electrons by impurities in semiconductors let us restrict to the case, when each impurity has one resonant energy level $\varepsilon_d$. In such a case the function $z(\varepsilon)$ in some interval of energies $\varepsilon_d - \Gamma < \varepsilon < \varepsilon_d + \Gamma$ of the width of $2\Gamma << \varepsilon_d$ describes the population of electron states localized at the impurity in resonance state. The parameter $\Gamma$ will be determined below. Off resonance at low energies ($\varepsilon < \varepsilon_r - \Gamma$) the localized contribution is absent, so that $z(\varepsilon) = 0$ and therefore one has

$$(1/\pi) \sum_l v_l \delta_l(\varepsilon) = 0, \qquad \varepsilon < \varepsilon_r - \Gamma \qquad (7)$$

On the other hand, at high energies ($\varepsilon > \varepsilon_r + \Gamma$) the localized part of electron density gives a contribution to в $n(\varepsilon)$, which corresponds to total population of resonance level. Then the quantity $z(\varepsilon)$ is equal to the multiplicity of degeneracy $v_r$ of the given level:

$$z(\varepsilon) = (1/\pi) \sum_l v_l \delta_l(\varepsilon) = v_r, \qquad \varepsilon > \varepsilon_r + \Gamma \qquad (8)$$

Eqns. (7) and (8) are the Friedel sum rules for the scattering phases. They reflect the fact that the number of the states for the motion with scattering without resonance is equal to the number of the states for free motion as far as the character of motion itself does not changed in the case of non-resonant scattering. On the other hand, in the presence of the resonance in the vicinity of the energy $\varepsilon_r$ the electron density involves the contributions of free motion and the state of localization. Accordingly in the resonance interval above the quantity $z(\varepsilon)$ is different from zero and from $v_r$ and dependent on the energy according to the behavior of the resonant phase of scattering $\delta_r(\varepsilon) = \pi z(\varepsilon)/v_r$. Let us emphasize that the application of Eqns. (7) and (8) together with (5) to considering the effects of resonance states rather than bound ones is the base of our approach. The contributions of bound states in the given scheme can be described by the phases multiple of the number $\pi$ and do not lead to non-trivial dependences on the energy.

For a more detailed treatment of the function $\delta_r(\varepsilon)$ let us note that according to the resonance scattering theory it can be presented as a sum of sharply changing resonant part, which is obtained from the expression (3), and slowly changing term $\delta_{sm}(\varepsilon)$ within the resonance interval. Let us write such a sum in the following form:

$$\delta_r(\varepsilon) = \pi/2 + arctg\,[(\varepsilon-\varepsilon_r)/\Delta] + \delta_{sm}(\varepsilon), \qquad (9)$$

The resonance contribution (the first two terms in (9)) describes the behavior of the function $\delta_r(\varepsilon)$ at $\varepsilon \to \varepsilon_r$:

$$\delta_r(\varepsilon) \approx \pi/2 + (\varepsilon-\varepsilon_r)/\Delta \qquad (10)$$

If the expression (10) is considered as linear approximation of the function $\delta_r(\varepsilon)$ in resonant interval, one has to let $2\Gamma = \pi\Delta$; then the resonance dependence (3) is described by rectangular peak and the term $\delta_{sm}(\varepsilon)$ is ignored. In order to describe in greater detail the behavior of $\delta_r(\varepsilon)$ the term $\delta_{sm}(\varepsilon)$ should be taken into account. Since the scale of changing the function $\delta_{sm}(\varepsilon)$ is large in comparison with $\Gamma$, it can be approximated by first terms of the expansion in Taylor series near $\varepsilon_r$. Because of the symmetric character of the function $z(\varepsilon)$ we have $z(\varepsilon_r) = v_r/2$, so that $\delta_{sm}(\varepsilon_r) = 0$; this equality follows from the form of writing the sum (9). Linear approximation for $\delta_{sm}(\varepsilon)$ is defined by the equality:

$$\delta_{sm}(\varepsilon) = (\varepsilon-\varepsilon_r)/\Delta_{sm}, \qquad (11)$$

in which the constant $\Delta_{sm}$ exceeds considerably $\Delta$ and $\Gamma$. Eqns. (7), (8) in such a approximation are fulfilled, as far as the following equation for determining $\Gamma$ is taken:

$$1/\Delta_{sm} = [\pi/2 - arctg(\Gamma/\Delta)]/\Gamma \approx \Delta/\Gamma^2 \qquad (12)$$

It follows from this equation that the width of resonant interval satisfies the inequalities $\Delta << \Gamma << \Delta_{sm}$. The parameter $\Delta_{sm}$ together with $\varepsilon_r$ and $\Delta$ characterizes the resonance at scattering by the given potential of an impurity. The limits of resonant interval within the scope of the approximation (11) are simulated by abrupt changes of the derivative of the function $z(\varepsilon)$ from zero to the values of the order of $1/\Delta_{sm}$ which are small as compared with resonant ones.

Let us consider now Eq. (6) and select the contribution of resonance phase from the sum in the right-hand side of Eq. (6) setting

$$\sum_l v_l\,\delta_l(\varepsilon) = v_r\,\delta_r(\varepsilon) + \sum_l{}' v_l\,\delta_l(\varepsilon) \qquad (13)$$

The prime at the sum sign denotes the elimination of the above contribution. Thereby the agreement about including the multiplicity of degeneration of the resonant level in $n_i$ remains in force. On the base of the above relations (9), (11), (12) one can write the following expression for the quantity $\delta_r(\varepsilon)$ different from zero in resonance interval only :

$$\delta_r(\varepsilon) = \pi/2 + arctg[(\varepsilon-\varepsilon_r)/\Delta] + [\pi/2 - arctg(\Gamma/\Delta)](\varepsilon-\varepsilon_r)/\Gamma \qquad (14)$$

In such a case the equality:

$$\sum_l{}' v_l\,\delta_l(\varepsilon) = 0 \qquad (15)$$

together with Eqns. (15), (16) is general formulation of Friedel's sum rule which in addition to Eqns. (7), (8) includes the resonance range. As will be seen below such a formulation allows us to use effectively the sum rule in studying the influence of resonance scattering on the conductivity.

2. Electron concentration and mean free path

Substituting the above relations into Eqn.(5) it should be taken into account that the function $n_e(\varepsilon)$, representing the contribution of free motion to the electron concentration, is slowly changed in the resonant interval and therefore close to the value $n_0 = n_e(\varepsilon_r)$. In linear approximation one has

$$n_e(\varepsilon) \approx n_0 [1 + (\varepsilon - \varepsilon_r)/\Delta_e], \quad \varepsilon_r - \Gamma < \varepsilon < \varepsilon_r + \Gamma, \quad (16)$$

where $\Delta_e = n_0 /[dn_e(\varepsilon_r)d\varepsilon_r]$. In this approximation according to Eqns. (5), (11), the function $n(\varepsilon)$ has the following form in the resonance interval:

$$n(\varepsilon) = n_0 + n_d [1/2 + (1/\pi) \, \text{arctg} \, \{(\varepsilon - \varepsilon_r)/\Delta\}] +$$
$$+ [n_d /\pi\Delta_{sm}) + (n_0/\Delta_e)](\varepsilon - \varepsilon_r), \quad (17)$$

where $n_d = n_i v_r$ is the concentration of donor electrons.

In calculating the conductivity we use the known expression for inverse mean free path $\Lambda^{-1}(\varepsilon)$ through the scattering phases, which can be written as follows:

$$\Lambda^{-1}(\varepsilon) = 2 n_i s_0(\varepsilon) \sum_l (l+1) \sin^2(\delta_l(\varepsilon) - \delta_{l+1}(\varepsilon)), \quad (18)$$

where $s_0(\varepsilon)$ is the coefficient independent of the scattering. Let us select in the sum over $l$ in this equality the terms involving the resonance phase:

$$(2r+1)[\sin^2 \delta_r - (1/2) \sin 2\delta_r \sin 2\varphi], \quad (19)$$

Thereby letting the non-resonant phases $\delta_l$ ($l \neq r$) be small and not taking into account the terms with $\sin^2 \delta_l$ we denote

$$\sin 2\varphi = (1+c_r)\sin 2\delta_{r+1} + (1 - c_r)\sin 2\delta_{r-1}, \quad (20)$$

where $c_r = 1/(2r + 1)$ and the argument $\varepsilon$, as in Eqn. (19), is omitted. Add now into Eqn. (19) in brackets the term $\sin^2 \varphi (1 - 2\sin^2 \delta_r)$. Equalize its non-resonant part by corresponding addition in Eqn. (18) and the part with the resonance phase is of the same order of magnitude as the terms which are not taken into account. Then it appears that the expression (19) can be replaced by $v_r \sin^2(\delta_r - \varphi)$ and in accepted approximation Eqn. (18) can be written in the form:

$$\Lambda^{-1}(\varepsilon) = n_d s_0(\varepsilon)[\alpha(\varepsilon) + \sin^2(\delta_r(\varepsilon) - \varphi(\varepsilon))] \quad (21)$$

The small quantities $\alpha(\varepsilon)$ and $\varphi(\varepsilon)$, entering Eqn. (21), characterize non-resonant phases. The function $\alpha(\varepsilon)$ describes the contribution of the all phases off resonance and $\varphi(\varepsilon)$ characterizes the difference of transport section of scattering from the total one.

In analyzing the dependence of the mean free path on the energy let us select first the resonance vicinity in which $|\varepsilon - \varepsilon_r| \ll \Gamma$. In this region the term $\delta_{sm}(\varepsilon)$ in resonance phase can be ignored and Eqn. (9) takes the form:

$$ctg\ \delta_r(\varepsilon) = (\varepsilon_r - \varepsilon)/\Delta \qquad (22)$$

Then taking into consideration that $\varphi(\varepsilon)$ is small in comparison with $\delta_r(\varepsilon)$ in the resonance region above we have:

$$\Lambda^{-1}(\varepsilon) = n_d\ s_0\ \{\alpha + 1/[1 + (\varepsilon_r - \varepsilon)^2/\Delta^2]\}, \qquad (23)$$

where $s_0 = s_0(\varepsilon_r)$, $\alpha = \alpha(\varepsilon_r)$. Off resonance interval according to Eqn. (21) the mean free path is equal to $(n_d\ s_0\ \alpha)^{-1}$. It follows from Eqn. (23) that at resonance $\Lambda(\varepsilon)$ falls off to the magnitude of the order of $(n_d\ s_0)^{-1}$ as $\varepsilon$ approaches $\varepsilon_r$. This manifests itself in concentration and temperature dependences of conductivity which are considered below. A necessary condition of existing the resonance effects is that the parameter $\alpha$ must be small as compared with unity.

Another specific energy region lies near the edges of resonance interval. In particular, the vicinity of upper edge, $\varepsilon = \varepsilon_r + \Gamma$, is of interest henceforth. Here the condition of applicability of Eqn. (22) is not fulfilled and in the resonance phase the contribution $\delta_{sm}(\varepsilon)$ should be taken into account. From Eqn. (14) one can obtain the following expression for small $\varepsilon_r + \Gamma - \varepsilon$:

$$(1/\pi)\ \delta_r(\varepsilon) = 1 - (\varepsilon_r + \Gamma - \varepsilon)/(\pi\ \Gamma_b), \qquad (24)$$

where $\Gamma_b = \Delta_{sm}/2 = \Gamma^2/2\Delta$. Further we substitute this expression into Eqn. (21) and believe that $\Gamma_b$ is considerably smaller than the specific scales of changing the quantities $\alpha(\varepsilon)$ and $\varphi(\varepsilon)$, and these quantities in the interval in question can be considered as constants. Thereby the contribution from $\varphi$ in combination $\varphi\ \Gamma_b$ we include in boundary energy $\varepsilon_r + \Gamma$, denoting it by $\varepsilon_b$. In this case Eqn. (21) takes the following form:

$$\Lambda^{-1}(\varepsilon) = n_d\ s_0\{\alpha + sin^2[(\varepsilon_b - \varepsilon)/\Gamma_b]\} \approx n_d\ s_0\ [\alpha + (\varepsilon_b - \varepsilon)^2/\Gamma_b^2] \qquad (25)$$

The energy dependence obtained describes the transition of the mean free path at the boundary to its non-resonant value.

3. Fermi energy

The results above allow us to analyze the concentration and temperature dependences. Consider first of all the electron concentration and the Fermi energy in ground state. It should be taken into account that the total concentration of electrons $n(\varepsilon)$ is given by the condition of electrical neutrality or the total number of electrons falling per unit volume of a system. In our system this number includes the electrons of resonance donors studied (concentration $n_d$) and conduction electrons arising from other donors and from the host (concentration $n_{0e}$). In metals, for which Friedel's theory has been developed [1], the concentration of conduction electrons does not depend practically on the donor concentration, while in a semiconductor in the case of interest to us the number of electrons is almost wholly defined by resonance donors. In both cases the following equality for ground state is valid:

$$n(\varepsilon_F) = n_d + n_{0e}, \qquad (26)$$

which is the equation for determining the Fermi energy $\varepsilon_F$ as a function of the donor concentration $n_d$. Under the conditions of resonance scattering of electrons the Fermi energy lies in resonance interval defined above and one should substitute Eqn. (17) for $n(\varepsilon_F)$ into Eqn.(26). It is not difficult to see that if the Fermi energy is near the resonance energy ($|\varepsilon_F - \varepsilon_r| \ll \Gamma$) and above this value, the contribution of the electron concentration dependence $n_e(\varepsilon)$ on the energy to $n(\varepsilon_F)$ is insignificant. For studying the dependence of the quantity itself $n_e(\varepsilon_F)$ on $n_d$ the Fermi energy value found should be substituted into Eqn. (16). In considering the concentration dependences of other quantities one can believe that after reaching the resonance the concentration of conduction electrons is equal to $n_0$. Thereby Eqn. (26) involves the difference $n_0 - n_{0e}$ which we shall denote below by $n_0$. Thus Eqn. (26) for the Fermi energy takes the following form:

$$(1/\pi)\, \delta_r(\varepsilon_F) = 1 - n_0/n_d \qquad (27)$$

From this equation in view of the condition $|\varepsilon_F - \varepsilon_r| \ll \Gamma$ according to Eqn. (22) one has:

$$\varepsilon_F - \varepsilon_r = \Delta\, ctg(\pi n_0/n_d) \qquad (28)$$

The resonance value of donor concentration $n_d$ is equal to $2n_0$. In increasing the concentration $n_d$ the Fermi energy is slowly increased remaining within the resonance interval. Using Eqn. (24) one obtains the expression for $\varepsilon_F$ near the boundary of interval $\varepsilon = \varepsilon_r + \Gamma$:

$$\varepsilon_F = \varepsilon_r + \Gamma - \pi\, \Gamma_b\, n_0/n_d \qquad (29)$$

### 4. Electron mobility

Let us apply now Eqns. (27) – (29) to describing the dependence of the electron mobility on the concentration of donor impurities in the resonance interval. This interval is bounded below by the value near $n_0$, so that with increasing $n_d$ the relation $n_0/n_d$ is changed from unity to small values. Let us consider the electron mobility in specific ranges discussed above in which Eqns. (28) and (29) are true. In the first of them the concentration $n_d$ takes the values near the resonance one, $2n_0$, and above in fulfilling the inequality $ctg(\pi n_0/n_d) \ll \Gamma/\Delta$. With using Eqns. (23) and (28) one obtains the following expression for the mobility $\mu$ in the given range:

$$\mu = \mu_0\, (n_0/n_d)[\alpha + \sin^2(\pi n_0/n_d)]^{-1} \qquad (30)$$

where $\mu_0$ is the mobility corresponding to the mean free path $1/n_0 s_0$. The effect of the resonance is in dropping the mobility from non-resonant magnitude $\mu_0(n_0/n_d)/\alpha$ to the values of the order of $\mu_0$. However the contribution of resonance phase $\sin^2(\pi n_0/n_d) \approx 1 - \pi^2 (1/2 - n_0/n_d)^2/2$ near the resonance is changed slowly therefore the mobility minimum at $n_d \approx 2n_0$ is highly smeared. With subsequent growth of the concentration $n_d$ the mobility increases, as far as the resonance phase contribution decreases until it reaches the values of the order of $\alpha$. As far as after that the mobility begins to drop because of non-resonant scattering, at the given values of $n_d$ its concentration maximum arises. This maximum is pronounced by virtue of the fact that the quantity $\alpha$ is small. It is clearly described by the expressions which are obtained from Eqn. (30) on condition that the relation $n_0/n_d$ is small:

$$\mu = \mu_0 (n_0/n_d)[\alpha + (n_0/n_d)^2]^{-1} = \mu_0 v/\pi (\sqrt{\alpha})(1+v^2)^{-1}, \qquad (31)$$

where $v = n_d (\sqrt{\alpha})/(\pi n_0)$. The maximum at $v = 1$ conforms to the values of $n_d$ inversely proportional to $\alpha$. Therefore for its description at small values of $\alpha$ the condition of applicability Eqns. (23) and (28) can be not fulfilled. However, it turns out that the expression for the mobility at large $n_0/n_d$, obtaining on the base of Eqns. (24), (29) coincides with (31). Consequently, one can believe that Eqn. (30) describes the concentration dependence of electron mobility in total resonance interval.

Maximum of the concentration dependence of electron mobility is one of basic effects of resonance scattering. It is in essence related to the stabilization of electron concentration in increasing the concentration of donor impurities in resonance region. In filling the localized states the effective charge of every impurity decreases, the electron scattering weakens and mobility increases. This increase goes on until non-resonance scattering becomes prevalent. This scattering characteristic of practically neutral impurities results again in the decrease of the mobility with increasing the concentration of scatters and as a result in the appearance of the mobility maximum.

The concentration maximum is reflected in the temperature dependence of electron mobility as well. Consider the ratio of the mobility $\mu(T)$ to its value discussed above at $T = 0$. The initial expression for this ratio is written in the following form:

$$\mu(T)/\mu = \int d\varepsilon (-\partial f/\partial \varepsilon)[\Lambda(\varepsilon)/\Lambda(\varepsilon_F)] = \int dE [4\kappa T ch^2(E/2\kappa T)]^{-1}[\Lambda(E)/\Lambda(0)], \qquad (32)$$

where $f$ is the Fermi function, $E = \varepsilon - \varepsilon_F$, $\kappa$ is Boltzman constant. In this expression the domain of integration is the resonance energy interval. As far as the concentration anomalies of temperature dependences are of interest to us, let us discuss firstly comparatively simple limit case of high concentrations of impurities, when the quantity $n_0/n_d$ is small and Eqns. (24), (29) are true. For this case Eqn. (32) takes the form:

$$\mu(T)/\mu = \int_{-\infty}^{\theta/T} dx [2ch(x/2)]^{-2}[1+v^2][(Tx/\theta - 1)^2 + v^2]^{-1}, \qquad (33)$$

where $\theta = \Gamma_b \sqrt{\alpha}/v\kappa$. The obtained relation describes the dependence which flattens in the limit of low temperatures and sharply drops above certain threshold temperature not always clearly definable. The scale of dropping is defined by the temperature $\theta$ and decreases with increase of impurity concentration. The threshold temperature is approximately defined as the highest between the quantities $\theta$ и $v\theta$, and so it is close to $\theta$ at lesser concentrations of impurities and to independent of $v$ temperature $v\theta$ at the larger ones. Thereby it is clear that at lesser concentrations the threshold is smearing (the scale of dropping and threshold temperature are of the same order), and with increase of the concentration it becomes more highly pronounced. The value $v = 1$ corresponding to the regarded above concentration maximum of the mobility is the limiting value of the appearance of threshold.

The like behavior is also described by the expressions for the mobility depending on temperature at the concentrations of impurities near the resonance value. In the conditions of the applicability of Eqns. (23) и (28) one can write the following expression for the function $\Lambda(E)$, entering the definition (32):

$$\Lambda(E)/\Lambda(0) = [1 + (1/\alpha)\sin^2(\pi n_0/n_d)]\{1 - (1/\alpha)[(E/\Delta + ctg(\pi n_0/n_d))^2 + 1 + 1/\alpha]^{-1}\} \quad (34)$$

The temperature dependences obtained by means of Eqn. (34) were applied to describing the experimental data in the work [2], in which the curves showing the effect of the concentration maximum of the mobility on the threshold temperature values are presented. It should be emphasized that such specific behaviors are in essence related to the fact of the stabilization of electron concentration in resonance interval which was discussed above.

5. Discussion of results and conclusion

The results above show that the application of the resonance scattering theory and Friedel's approach to the system of electrons scattering by donor impurities in a semiconductor allows us to predict the concentration maximum and its related specific behavior of temperature dependences of electron mobility. Such anomalies were revealed experimentally and studied in detail in mercury selenide with iron impurities. The interpretation of experimental data on the basis of the approach developed by us is presented in the work [2]. Up to now in the investigations on this problem (see Refs. [3,4]) the observed maximum of the mobility is interpreted on the basis of another approach which was formulated by J.Mycielski [5]. It admitted simultaneous existence of ionized and non-ionized donor states at the resonance $\varepsilon_F = \varepsilon_d$. Such an assumption means actually that the state of an electron at the donor is bound similarly to the states in forbidden band of crystal rather than resonant one. Using the given approach for describing the observed maximum of the mobility the authors substantiated additionally with the help of model calculations the existence of the ordering of ionized donors [5, 6] and introduced other assumptions about the structure of impurity system. As a whole the analysis detailed allows us to make sure that such an interpretation of the mobility maximum involves the model assumptions which are not rigorously justified. At the same time it appears that the experimental facts conform most likely to the idea about resonant (in typical sense of this term of quantum-mechanical scattering theory) state of an electron on the impurity and identical partially ionized state of all the donors respectively. It is such an approach that is presented in our paper. It allows us to explain the behavior of the mobility not assuming the ordering of donors, the existence of which can be unlikely accepted as rigorously proved in the HgSe:Fe crystals at the moment. But of course, in order to justify rigorously the validity of one of two approaches for some system it would be well to appeal to the data of such experiments in which the effects of resonance and bound states would be uniquely distinguishable. The complexity of this problem is that for qualitative interpretation of a number of experiments this distinction is of no importance and the researchers do not take into account it. Therefore for the present it is difficult to apply such a method to the HgSe:Fe system as far as in experimental investigations carried out the similar problems were practically not stated.

Thus we have shown that the scattering of electrons by donor impurities having the resonance energy level in conduction band of semiconductor leads to the stabilization of electron concentration and to maximum of their mobility depending on the impurity concentration and also to specific anomalies of the mobility depending on the temperature. There is reason to believe that the theory, which predicts these effects in the framework of generalized Friedel's approach, can serve as dependable basis for the interpretation of experimental data.

The work was supported by RFBR, Grant No 03-02-16246.